# Control of coherent information via on chip photonic-phononic emitter-receivers

Heedeuk Shin[1], Jonathan A. Cox[2], Robert Jarecki[2], Andrew Starbuck[2], Zheng Wang[3], and Peter T. Rakich[1*]

[1] Department of Applied Physics, Yale University, New Haven, CT 06520, USA

[2] Sandia National Laboratories, PO Box 5800, Albuquerque, NM 87185, USA

[3] Department of Electrical and Computer Engineering, University of Texas at Austin, Austin, TX 78758, USA

[*] email: peter.rakich@yale.edu

**Rapid progress in silicon photonics has fostered numerous chip-scale sensing, computing, and signal processing technologies. However, many crucial filtering and signal delay operations are difficult to perform with all-optical devices. Unlike photons propagating at luminal speeds, GHz-acoustic phonons with slow velocity allow information to be stored, filtered, and delayed over comparatively smaller length-scales with remarkable fidelity. Hence, controllable and efficient coupling between coherent photons and phonons enables new signal processing technologies that greatly enhance the performance and potential impact of silicon photonics. Here, we demonstrate a novel mechanism for coherent information processing based on traveling-wave photon-phonon transduction, which achieves a phonon emit-and-receive process between distinct nanophotonic waveguides. Using this device physics—which can support 1-20GHz frequencies—we create wavelength-insensitive radio-frequency photonic filters with an unrivaled combination of stopband attenuation, selectivity, linewidth, and power-handling in silicon. More generally, this emit-receive concept is the impetus for numerous powerful new signal processing schemes.**

A variety of micro- and nano-scale systems have recently been used to enhance and tailor the interaction between photons and phonons through engineerable structures and materials[1–12]. In the quantum limit, highly controlled photon-phonon couplings permit ground state cooling of microscopic optomechanical systems[13–16], and offer a unique quantum interface between the optical, microwave, and phononic domains[17,18]. In the classical limit, such photon-phonon interactions are also of significant technological value since they enable new chip-scale radio-frequency (RF) photonic signal processing approaches that combine the merits of both photonic and phononic excitations[3,19,20]. Specifically, with efficient and reversible conversion between photonic and phononic domains, long coherence time and low velocity GHz acoustic phonons can be used to slow, store, and frequency-shift optical signals[4,5,10,21–23]. By harnessing such interactions, phonons can act as coherent memory[21,22], produce filters with ultra-high roll-off[9,24], and enable a host of processes having no analogue in all-optical silicon photonics[3,5,20,23,25–27]. While engineerable coupling between resonant photonic and phononic modes has been achieved in a variety of chip-scale systems[3–6,13–16,19,20,28–31]a significant challenge to the creation of robust new silicon-based coherent signal processing technologies remains: the realization of narrow-band filters that simultaneously achieve high optical power handling, low signal distortion, and wavelength insensitivity in silicon[32,33].

We address this challenge through a novel chip-scale multiport photonic-phononic emitter-receiver (PPER) system that produces strong photon-phonon coupling *without* use of optical resonance. This traveling-wave geometry enables independent control of guided photonic and phononic modes, as well as their interaction. In this system (Fig.1), forces produced by traveling optical signals transduce coherent phononic signals in the core of an (emitter) optical waveguide; the surrounding phononic crystal superstructure then shapes the transfer of the phononic signal to an adjacent (receiver) optical waveguide, which converts the signal back to the optical domain through photoelastic coupling (see Fig. 1b). The phononic crystal superstructure provides tailorability of the phononic transfer function, permitting control of transduction bandwidth, frequency, and conversion efficiency over a range of GHz frequencies (see Fig. 1c). Owing to the traveling-wave nature of this two-port PPER system, information transfer is independent of the wavelength and temporal coherence of the optical waves in the emit- and receive-ports. As a result, broadband wavelength conversion, radio frequency signal mixing, and ultra-narrow-band signal filtering are achieved together with high optical power handling (>100 mW), a much needed combination in silicon photonics[32,33].

To date, much of silicon photonics has relied on optical resonators for RF-photonic filtering operations. However, since the RF signal bandwidths and the optical carriers differ in frequency by many orders of magnitude, narrow band RF filtering becomes a very challenging task. For instance, a 2nd order filter with 3-MHz bandwidth—comparable in performance to this PPER system—would require two ultra-high Q-factor ($\sim 10^8$) optical cavities with precisely controlled optical coupling and frequency alignment between them. Nonlinear absorption in silicon severely limits the power levels in such high-Q cavities[34]. Moreover, narrow linewidth lasers must be actively frequency-stabilized relative to the cavity resonances when used as filters. These factors often limit the practicality of narrow-band all-optical filtering operations in silicon RF-photonics. Alternatively, backwards stimulated Brillouin scattering (backward-SBS) has been studied as a means of implementing RF-photonic filters[9,24]. However, such backward-SBS based filtering schemes have proven difficult to implement in silicon.

More generally, the physics of this new PPER system provides a number of advantages for chip-scale signal processing. Since the emit-receive functionality is inherently wavelength insensitive, and the spatially separate ports of PPER system produce negligible optical cross-talk, optical wavelength conversion is readily achieved through the emit-receive process. Information can be transduced between any two wavelengths in the emit- and receive-ports provided that the optical group-velocity walk-off of the propagating signals is small compared with the signal modulation period. (See the supplementary information.) Hence, the wavelengths of the emit- and receive-ports are unconstrained; both narrow band filtering and wideband wavelength conversion can be achieved in this system. More importantly, independent control of photon and phonon modes opens up a large design space for a range of new coherent signal processing schemes in the context of silicon photonics.

In what follows, we first examine the physics of strong photon-phonon coupling in a Brillouin-active waveguide, which serves as a key building block for wavelength-insensitive 2[nd] order RF-photonic filters (later discussed in this Article). This waveguide system utilizes phononic crystal defect states to produce strong photon-phonon coupling at selected frequencies in the phononic crystal stopband. Building on this device physics, two such phononic crystal defect states are evanescently coupled to produce the traveling-wave PPER system seen in Fig 2i. Adjusting the device geometry, we show that the phononic crystal supermodes of the system can be engineered to shape the signal transfer response of the emitter-receiver pair. Applying this phonon emitter-receiver physics, we synthesize high performance 2-port RF-photonic filters.

**Brillouin-active phononic crystal waveguide.** We first examine the strength of traveling-wave photon-phonon transduction generated by a stand-alone silicon waveguide embedded within a silicon nitride membrane (see Fig. 2a). This structure guides both optical photons and acoustic phonons while the photoelastic response of the silicon waveguide core mediates reciprocal photon-phonon coupling. As the active component of the emitter-receiver pair seen in Fig. 1, the transduction efficiency of this system is of critical importance. The highly tensile silicon nitride membrane is patterned to form two phononic crystal regions, consisting of a square lattice of holes, placed symmetrically about the silicon waveguide core. Bragg reflection produced by these phononic crystal regions creates guided phonon modes (i.e., defect modes) in the phononic stopband. The silicon waveguide core centred within the phononic waveguide yields tight confinement of the guided optical mode through total internal reflection. A top-down SEM image of the fabricated Brillouin-active membrane waveguide with a surrounding phononic crystal structure (PnC-BAM waveguide) is seen in Fig. 2b along with a magnified view of the waveguide cross-section ($950 \times 220 \text{ nm}^2$) in Fig. 2c. Throughout this article, the lattice constant, $a_o$, is 1 μm, and the radius of holes, $r_o$, is 0.385 μm. In comparison to a related structure[35], the enlarged silicon core of this waveguide results in lower propagation loss ($< 1 \text{ dB/cm}$), lower nonlinear absorption[36], and higher power handling ($> 110 \text{ mW}$), which contribute to significant performance enhancements. For further details about power handling, see the supplementary information.

Strong coupling between these co-located optical and phononic modes is mediated by optical forces generated within the silicon waveguide core[37]. This form of traveling-wave photon-phonon coupling is termed forward-stimulated Brillouin scattering (forward-SBS) or stimulated Raman-like scattering[38,39]. Through forward-SBS, energy can be transferred between optical pump and signal waves propagating within the waveguide. Full-vectorial multi-physics simulations[35,37,40] were used to compute photon-phonon coupling within a PnC-BAM waveguide of dimension $W_o = 7.2 \text{ μm}$, and reveal that the PnC defect mode of Fig. 2d is efficiently excited at a frequency of 3.72 GHz within the stopband in Fig. 2g. Figure 2e and 2f show the computed $E_x$-field of the fundamental TE-like mode of the silicon waveguide and the $x$-component of the corresponding electrostrictive force density, respectively. A vector phase matching diagram for forward-SBS (yellow arrows) is shown in Fig. 2a. Here, a phonon of wave-vector $\mathbf{K} = \mathbf{k}_1 - \mathbf{k}_2$ and frequency $\Omega = \omega_1 - \omega_2$ mediates this interaction, where $\mathbf{k}_j$ and $\omega_j$ are the wave-vector and frequency of interacting optical waves. Phase matching requires that the group velocity of the optical signal match the phase velocity of the guided phonon mode ($\Omega/\mathbf{K}$). This condition is only satisfied by guided phonon modes[35] with ultra-slow group velocity of ($\partial\Omega/\partial|\mathbf{K}| \sim 1 \text{ m s}^{-1}$).

The computed phononic dispersion relation for the fabricated 2D phononic crystal structure is seen in Fig 2g[41]. We limit our attention to the part of the dispersion diagram with $k_y \sim 0$, commensurate with the phononic modes participating in forward-SBS[35]. As seen from the highlighted (yellow) regions of Fig 2g, a phononic stopband extends from $2.6 - 4.5$ GHz. By designing the dimension of the phononic defect, $W_o$, individual Brillouin-active PnC defect modes can be created at select frequencies within this stopband, yielding strong Brillouin resonances and strong photon-phonon coupling.

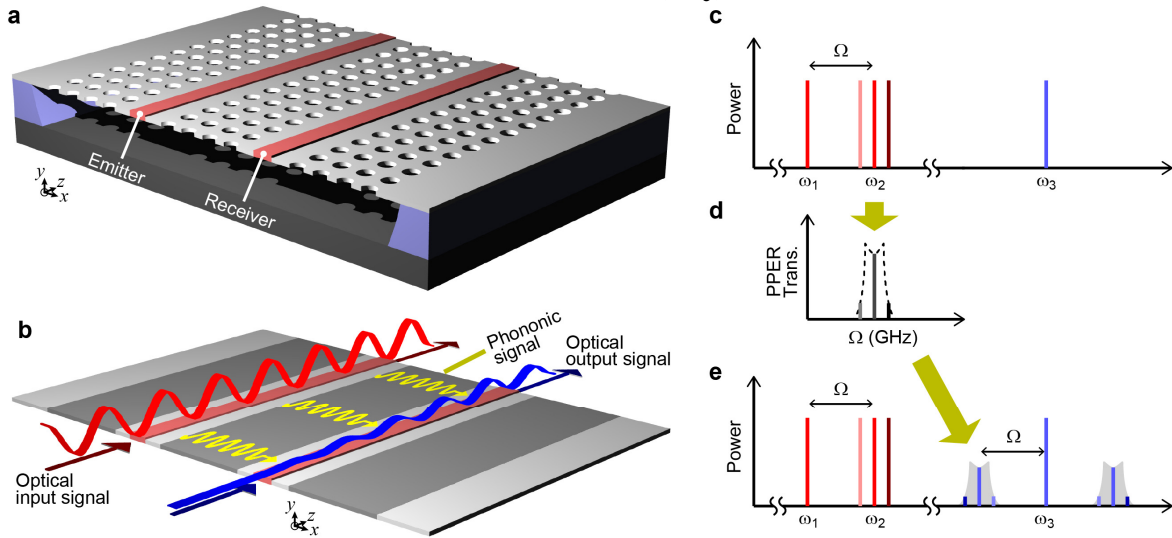

**Figure 1| Traveling-wave photonic-phononic emitter-receiver (PPER). a,** Schematic of a PPER system consisting of two silicon optical waveguides (red) embedded in a phononic crystal membrane (grey). **b,** Diagram showing principle of PPER operation. Red, blue, and yellow curves are the optical input signal, optical output signal, and transduced phonon waves, respectively. Information is encoded on the red wave (emitter) through amplitude modulation; transduced phonons then couple this information to a monochromatic blue wave (receiver) of disparate wavelength via parametric coupling. **c-e,** Characteristic spectra showing the input (**c**) and output (**e**) optical signals and the response produced by phononic supermodes that mediate information transfer (**d**). In the emitter port, a pump field ($\omega_2 = \omega_1 + \Omega$) is swept relative to a local oscillator ($\omega_1$) to produce an amplitude modulated beat-note. Optical forces generated in the emitter waveguide drive the excitation of phonons. Information is transferred between the emitter and receiver ports by phonon supermodes that produce the transfer function in **d** (black dashed). Information is encoded on the blue probe field ($\omega_3$) of the receiver port for frequencies within the transfer function of the phononic supermodes (shaded gray in **e**).

The photon-phonon coupling was quantified through studies of the fabricated PnC-BAM waveguide (Fig. 2b-c), which is suspended continuously over a 4-mm length. Lithographically tapered silicon input and output couplers ensure that only the fundamental TE-like mode (seen in Fig. 2e) is excited. The strength of photon-phonon coupling was experimentally determined through heterodyne four-wave mixing experiments[35], enabling the study of Brillouin active phonon modes between $0.5 - 9$ GHz. The dashed inset of Fig. 2h shows the result of one such four-wave mixing measurement; coherent interference between the Kerr and Brillouin nonlinear susceptibilities produce a characteristic Fano-like line-shape. The analysis of this line-shape provides a measurement of Brillouin nonlinearity relative to the intrinsic Kerr nonlinearity[35], yielding a Brillouin gain coefficient of $G_{SBS} = 2|\gamma_{SBS}| = 1960 \pm 355$ W$^{-1}$m$^{-1}$ and a linewidth of 1.2 MHz, (or Q-factor of $\sim 3160$ at a centre frequency of 3.72 GHz). Note that this geometry enables the synthesis of isolated Brillouin-active PnC defect modes within the phononic crystal stopband (yellow regions), simultaneously producing strong photon-phonon coupling, high power handling, and low propagation losses.

**Physics of the phonon emitter-receiver.** In addition, such phononic crystal geometries permit control of phonon emission and coupling as the basis for new traveling-wave phonon emit-receive functionalities. The anatomy of a traveling-wave phonon emitter-receiver pair—in the form of a dual PnC-BAM waveguide—is shown in Fig. 2i for comparison with the top-down SEM image of Fig. 2j. This system is comprised of two silicon waveguides embedded within a silicon nitride phononic crystal superstructure having two PnC line-defects ( $W_0 = 5.7$ μm ). Each PnC line-defect is bounded by symmetrically placed PnC regions ($N$ periods each), while a centrally located PnC region ($N_c$ periods) separates the line defects by a centre-to-centre distance of $[(N_c - 1) \times a_0 + W_0]$. Figures 2i-j show the special case of $N = N_c = 6$. Silicon waveguides are centred in each line-defect (as seen in Fig. 2j), producing a dual channel system with mirror symmetry.

The diagram of Fig. 2k illustrates the basic physical principle by which the phonon emit-receive geometry transfers information from light within emitter waveguide (Wg-A) to light in receiver waveguide (Wg-B). Through this reciprocal process, electrostrictive optical forces produced by optical signals within Wg-A drive the excitation of coherent phonons; the surrounding phononic crystal superstructure then shapes the transfer of phononic information between Wg-A and Wg-B. Through this process, phononic energy transfer is mediated by phononic crystal supermodes consisting of evanescently coupled PnC defect modes within the superstructure (as illustrated by Fig. 2m). The transduced phononic information is then encoded on optical waves carried by Wg-B through photoelastic coupling. In contrast to the phononic properties of the system, negligible optical cross-talk occurs between Wg-A and Wg-B since the guided optical modes decay rapidly ($\sim 60$ nm) outside of the silicon waveguide core (as seen in Fig. 2e). The coupling rate ($\mu$) between the phononic defect modes is mediated by the central PnC coupling region (of $N_c$ periods), while the external decay rate ($\tau_e^{-1}$) from each defect mode is determined by the PnC cladding region (of $N$ periods) on either side of the device.

The spatial dynamics of such phonon-mediated coupling between the optical waves can be treated analytically using temporal coupled-mode theory[42,43] (See the supplementary information). As illustrated by Fig. 2k, optical fields $E_1^a(\omega_1, t)$, $E_2^a(\omega_2, t)$, and $E_3^b(\omega_3, t)$ are injected into the system, and we seek the parametrically generated signal $E_s^b(\omega_s, t)$ at the output of Wg-B. Optical forces produced by the interference between $E_1^a$ and $E_2^a$ drive the phonon supermodes; these can be expressed as a linear combination of the elastic displacement fields $e_a(x, y)$ and $e_b(x, y)$, representing the phononic crystal defect modes in Wg-A and Wg-B, respectively (as seen in Fig 2n). The modal hybridization, and the resonant transduction between the Brillouin active modes are accurately captured in terms of the modal coupling rate ($\mu$) the net modal decay rate ($\tau_{net}^{-1}$). Using optical forces to source the excitation of the PnC defect mode in Wg-A, modal perturbation theory can be used to determine the parametric growth of $E_s^b$ in Wg-B. The growth rate of the signal wave amplitude, $B_s$, in the $z$-direction of propagation is,

$$\frac{\partial B_s}{\partial z} = i\left[\frac{\omega_3 \tau_{net}}{2\Omega_0} \frac{\langle e_a, f_n^b \rangle \langle f_n^a, e_b \rangle}{\langle e_a, \rho e_a \rangle} \frac{2\mu/\tau_{net}}{\Gamma_-(\Omega)\,\Gamma_+(\Omega)}\right] A_1 A_2^* B_3 \quad (1)$$

$$= i[\gamma_{a \to b}(\Omega)]\, A_1 A_2^* B_3.$$

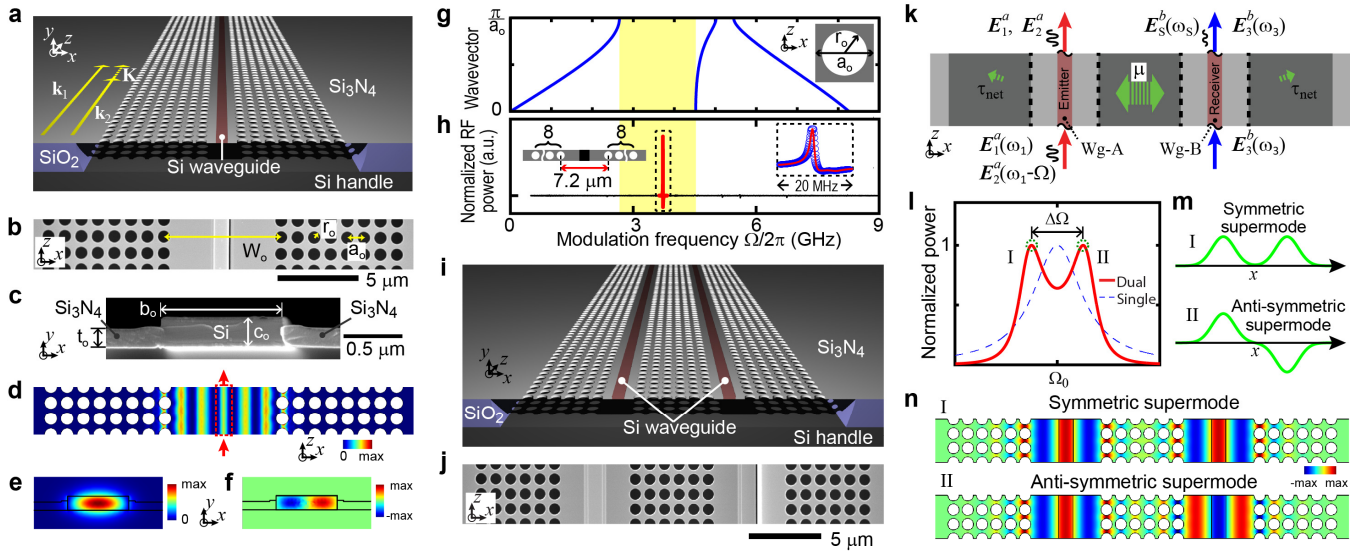

**Figure 2 | Brillouin-active membrane waveguide embedded in phononic crystal (PnC) structures. a**, Schematic showing the anatomy of PnC-BAM waveguide with the wave-vectors of the optical pump ($k_1$), the scattered light ($k_2$), and the phonon ($K$). **b**, Top-down SEM image of a segment of the fabricated PnC-BAM waveguide. The defect size of the PnC structure, $W_0$, is defined as the centre-to-centre distance between inclusions on either side of silicon waveguide. The lattice constant, $a_0$, is 1 μm, and the radius of holes, $r_0$, was $0.385a_0$. **c**, High resolution SEM image of the cross-section of the silicon waveguide core within the nitride membrane. The size ($b_0 \times c_0$) of the silicon waveguide is 950 × 220 nm$^2$, and the thickness ($t_0$) of silicon nitride is 130 nm. **d**, Computed acoustic energy density in a unit cell of PnC-BAM waveguide at a resonant frequency of 3.72 GHz. The red dotted box indicates the position silicon waveguide core. **e**, Computed $E_x$ fields of the guided optical mode in a silicon waveguide with a size of 950×220 nm$^2$. **f**, Computed $x$ component of the electrostrictive force density from the guided optical mode. **g**, Bulk dispersion diagram of the 3D phononic crystal cladding (unit cell shown in the inset) showing phonon frequency versus $x$-direction phonon wave vector. **h**, Spectrum of Brillouin responses obtained through heterodyne four-wave mixing measurements. The output signal from the PnC-BAM waveguide for $W_0 = 7.2$ μm is normalized to that of a reference silicon waveguide (Brillouin inactive) under identical experimental conditions. The left inset is the schematic geometry of the PnC-BAM waveguide. **i**, Schematic showing dual-channel PnC-BAM waveguide that forms the PPER system under study. **j**, Top-down SEM image of a portion of the dual-channel PPER system. **k**, Pictorial representation of PPER systems. $\mu$ and $\tau_{net}$ represent the coupling rate between adjacent PnC defect modes and net phonon decay rate, respectively. **l**, Sketch showing the parametric responses for single (blue dashed) and dual (red solid) channel PnC-BAM waveguides. **m**, Illustration of the displacement distribution for symmetric (I) and anti-symmetric (II) supermodes. **n**, Simulated elastic displacement fields associated with symmetric and antisymmetric supermodes of the PPER system.

Here, $\gamma_{a \to b}(\Omega)$ represents the phonon mediated coherent coupling from Wg-A to Wg-B, and $\Gamma_\pm(\Omega) \equiv [\Omega - (\Omega_0 \pm \mu) + i/\tau_{net}]$. We use the following definitions: $\Omega_0$ is the natural frequency of uncoupled phonon modes; $\rho(x, y)$ elastic medium mass density; $\tau_{net}^{-1}$, $\tau_e^{-1}$, and $\tau_o^{-1}$ are the net, external, and internal phonon decay rates, where $\tau_{net}^{-1} = \tau_e^{-1} + \tau_o^{-1}$; $P_j^a$ and $P_j^b$ are the powers carried by $E_j^a$ and $E_j^b$; $A_j$ and $B_j$ are the normalized wave amplitudes of $E_j^a$ and $E_j^b$ such that $|A_j|^2 = P_j^a$ and $|B_j|^2 = P_j^b$; and $f_n^a(x, y)$ and $f_n^b(x, y)$ are the power normalized force densities produced by light in Wg-A and Wg-B under continuous-wave excitation.

As seen in Fig. 2l, phonon mediated coherent coupling from Wg-A to Wg-B, or $\gamma_{a \to b}(\Omega)$, exhibits a sharp $2^{nd}$ order response with poles at $\Omega_\pm = \Omega_0 \pm \sqrt{\mu^2 - 1/\tau_{net}^2}$. These resonances correspond to symmetric and anti-symmetric phononic supermodes as seen in Fig. 2m-n. Note the second order response produced by this doubly resonant system produces far sharper roll-off than the first order (Lorentzian) response of single channel systems, as illustrated in Fig. 2l. Due to the symmetry of this geometry, the elastic displacement field ($e_j$), the power-normalized force density ($f_n^j$), and the overlap $\langle e_j, f_n^j \rangle$ are effectively identical in both waveguides ($j = a, b$). As a consequence, $|\gamma_{a \to b}(\Omega_\pm)| = G_o/2$, where $G_o$ is the single-waveguide Brillouin gain (of Ref. [40]) in the limit as $\tau_o^{-1} = 0$.

In the limiting case when $\mu = 0$ ($N_c = \infty$), we see that no information can be transduced from Wg-A to Wg-B, and the phononic eigenmodes of these waveguides are degenerate (see the dashed curve of Fig. 2l). However, for finite couplings ($\mu > 0$), hybridisation of the Brillouin-active phonon modes produce both symmetric and anti-symmetric supermodes (see Fig. 2m-n) with resonant frequencies (or poles) $\Omega_+$ and $\Omega_-$, respectively. These hybridized phonon modes mediate the transfer of coherent information between the two optical waveguides. In experiments, we explore the low gain regime, where the signal power is given by $P_s^b = |\gamma_{a \to b}(\Omega)|^2 P_1^a P_2^a P_3^b L^2$. As a result, the power of the signal wave ($P_s^b$) generated over a length $L$, increases quadratically with pump power ($\propto P_1^a P_2^a$) and length ($\propto L^2$). Note that the concise analytical model (summarized by Eq. 1) describes the physics of this multi-port PPER system with a few basic physical parameters, greatly reducing the complexity of this nontrivial 3D system. Most crucially, this model provides a means of understanding spectral response of the PPER system and its overall efficiency (as discussed above).

## Demonstration of a phonon emitter-receiver.

The photonic-phononic emit-receive functionality is demonstrated using the fabricated dual channel PPER of Fig. 2i. In this system, information encoded on pump-waves $E_1^a(\omega_1)$ and $E_2^a(\omega_2)$ at a wavelength of 1547 nm generates time-varying optical forces in Wg-A; these forces drive the excitation of hybridized phonon supermodes that transfer information from Wg-A to Wg-B. Through this process, the excited phonons coherently produce a signal wave, $E_s^b(\omega_s)$, through traveling-wave phase modulation of the injected probe wave, $E_3^b(\omega_3)$, at wavelength of 1536 nm in Wg-B. The transduced signal is then measured at the output of Wg-B through heterodyne detection, using the apparatus seen in Fig. 3. Since Wg-A and Wg-B are optically decoupled, optical crosstalk is negligible. As seen in Fig. 3, the frequency of excited phonon is controlled by changing the RF modulation frequency ($\Omega$), enabling quantitative study of the dual-channel PPER response from $1 - 9$ GHz. For further details, see Experimental Methods. In this way, this dual channel system behaves as a high fidelity PPER pair.

The measured response of a fabricated emitter-receiver ($W_o = 5.7$ μm, $N = N_c = 6$), with a 7-mm interaction length, is seen in Fig 3b. A sharp second-order frequency response is centred at 2.93 GHz (blue), demonstrating efficient phonon mediated information transfer between Wg-A and Wg-B. The measured data (blue) show a full width at half maximum of 3.15 MHz, corresponding to an aggregate Q-factor of ~930. Comparing to the theoretical second-order transfer function (red) from equation 1, with $\mu = 8$ MHz, $1/\tau_{net} = 6$ MHz, and $Q_{net} = \Omega_o \tau_{net}/2$ ~1530, we find excellent agreement, which suggests remarkable structural homogeneity and low phonon dissipation over the entire device length. This sharp $2^{nd}$ order response is highly desirable as it yields high selectivity against unwanted signals (or noise). Moreover, building on such device

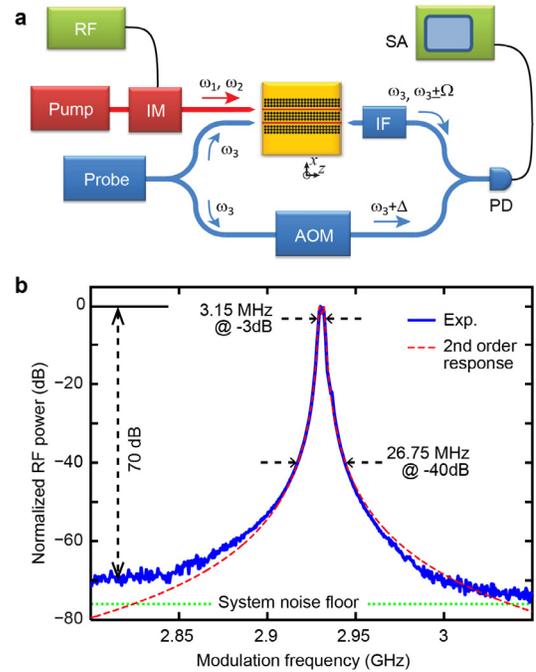

**Figure 3| RF photonic response of a dual channel PnC-BAM waveguide.** a, Schematic diagram of the apparatus used to measure the Brillouin nonlinearities of PnC-BAM waveguides. Pump: 1547-nm laser, Probe: 1536-nm laser, IM: intensity modulator, RF: RF generator for the intensity modulator, AOM: Acousto-optical modulator, IF: interference filter, PD: fast speed photodiode receivers, and SA: spectrum analyser. b, Normalized RF response of a PPER system. The theoretical prediction with second-order response (red dashed) is atop the experimental data (blue). System noise floor is expressed as a dotted green line.

topologies, a variety of higher order responses can be created using cascaded structures[43,44].

As seen in Fig. 4b, the measured 2nd order response (blue) shows a tremendous out-of-band rejection of > 70 dB; optical crosstalk poses no limitation to the dynamic range of the measured frequency response. The contrast of these measurements is limited only by our measurement noise floor (green dotted). Hence, to an excellent degree, information is transduced between Wg-A and Wg-B solely by the engineered phonon supermodes of the system. Through the measurements seen in Fig. 4, fibre-to-waveguide coupling efficiency limits the pump wave powers ($P_1^a$, $P_2^a \approx 3.5$ mW) in device, yielding a peak signal generation efficiency ($P_s^b/P_3^b$) of $\sim 10^{-4}$. Note that this efficiency can be significantly enhanced by increasing pump power ($P_s^b \propto P_1^a \cdot P_2^a$), interaction length ($P_s^b \propto L^2$), and Brillouin gain ($P_s^b \propto G_o^2$). For instance, with pump powers of $P_1^a = P_2^a \approx 70$ mW and longer interaction lengths ($L \approx 1.4$ cm) efficiencies of greater than 10 percent are readily achievable. Moreover, significant enhancements in the photon-phonon coupling strength ($G_o$) can be realized by modifying the geometry of traveling-wave phonon emit-receive structures.

From these measurements, we demonstrate a wavelength-insensitive transfer function with an unrivaled combination of stopband attenuation, selectivity, linewidth, and power-handling in silicon photonics. These characteristics are essential for many high-performance applications. Viewing this system as an RF-photonic filtering technology, we see that PPER pair has a desirable shape factor (8.5), as well as remarkably high slope ($\sim$20 dB/3.55 MHz)[45].

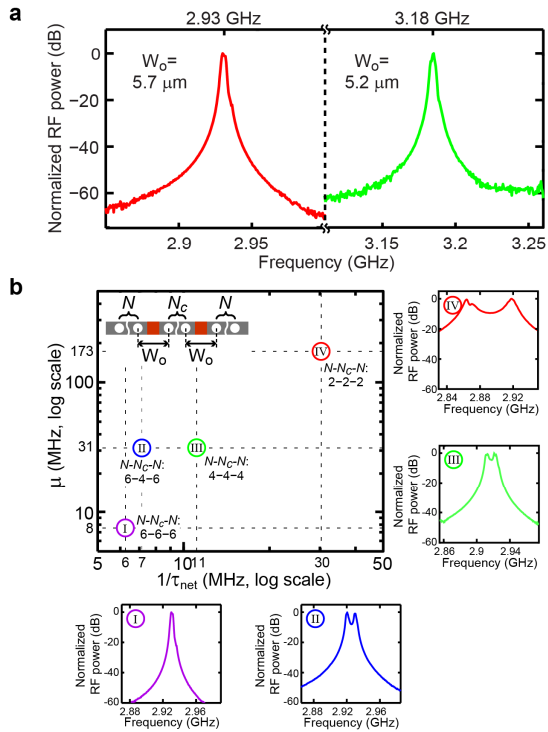

**Figure 4| Engineered response of dual-channel PPER. a,** Measured RF responses of PPER systems for $W_o = 5.7$ μm (red) and $W_o = 5.2$ μm (green) versus frequency. **b,** Coupling rates (μ) versus net decay rates ($\tau_{net}^{-1}$) extracted from measured RF responses (I, II, III, and IV) of PPER systems with $N - N_c - N = [6\text{-}6\text{-}6, 6\text{-}4\text{-}6, 4\text{-}4\text{-}4,$ and $2\text{-}2\text{-}2]$, respectively. Results are shown in log-log scale. The inset in the top-left corner is schematic geometry of the BAM waveguides. The insets in the bottom and right corners are the normalized RF power responses (I, II, III, and IV) corresponding to the combinations of the PnC layer numbers of $N - N_c - N = [6\text{-}6\text{-}6, 6\text{-}4\text{-}6, 4\text{-}4\text{-}4,$ and $2\text{-}2\text{-}2]$, respectively.

(The shape factor is the ratio of full width at -40 dB to full width at -3 dB.) Moreover, the rejection bandwidth of > 65 dB was observed over 1.8-GHz stopband (2.6 − 4.5 GHz). See the supplementary information for the response spectra over stopband range. Note that all of these characteristics are controllable by engineering the phononic crystal structure.

**Phononic supermode engineering.** As described above, the PPER response is determined solely by phonon supermodes straddling both waveguides. These supermodes are controllable by engineering the phononic crystal structure. Here, we experimentally demonstrate the ability to engineer the response of phonon emitter-receiver systems. Note that the centre frequencies of PPER supermodes can be tailored by engineering the defect size $W_o$. In Fig. 5a, the emit-receive responses for two different defect sizes are shown for the number of phononic crystal layers, $N = N_c = 6$. While the lineshape does not change significantly, the centre frequency is shifted 250 MHz as the defect size is lithographically varied from $W_o = 5.7$ μm (red) to 5.2 μm (green). Hence, we can readily tailor the centre frequency of information transduction while preserving the filter shape.

In addition, the PPER transfer function can be shaped by lithographically controlling frequency splitting and decay rate of the phononic crystal supermodes. Hence, phononic supermodes can be engineered by varying, $N$, $N_c$, and $W_o$. To demonstrate this control, we examined the RF response of PPER systems with a fixed $W_o$ of 5.7 μm, while lithographically varying $N$ and $N_c$ numbers. Experimentally measured responses for PPER devices with $N - N_c - N$ values of [6-6-6, 6-4-6, 4-4-4, and 2-2-2] are seen in Fig. 4. From these data, the coupling rate (μ) and the net decay rate ($1/\tau_{net}$) are extracted by fitting the analytical response function (equation 1) to the experimental data of RF response of each device (insets of Fig.5b). As seen in Fig. 5b, the coupling rate and the net decay rate can be tailored over a significant range by varying the numbers of the PnC coupling layers, $N_c$, and the PnC cladding layers, $N$, respectively. Note that the phononic coupling rate between Wg-A and Wg-B increases with smaller $N_c$ values while the phonon lifetime drops with smaller $N$ values. Hence, this topology provides unprecedented control over the centre frequency, bandwidth, shape-factor, and slope.

## Discussion

In this Article, we have demonstrated the traveling-wave photonic-phononic emitter-receiver and studied the frequency response of coherent information transduction through such device physics. As an RF filter, this PPER system simultaneously possesses high dynamic range (70 dB), high Q-factor, wide rejection bandwidth (~1.8 GHz), and high selectivity (bandwidth of 3 MHz, low shape factor of 5, and slope of 20/3 [dB/MHz]). The underlying phonon-mediated mechanism could form the basis for a host of powerful new coherent information processing technologies involving wavelength conversion, amplifier, RF mixing, and RF photonic filter. While we have limited our demonstration to classical signal processing in this paper, one can readily extend this mechanism to single-photon-level probe light with negligible optical crosstalk, enabling forms of single-photon frequency conversion[46] and quantum information processing[47].

More generally, this compound emit-receive system behaves as a 2-port optical system with negligible optical cross-talk and back-reflection: information is transferred from one port to another through phononic information transduction. Since this traveling-wave (or reflectionless) geometry negates the need for optical isolators, this platform is directly compatible with silicon-photonic systems. Moreover, the relaxed phase-matching conditions of this forward-scattering geometry enable narrow-band phononic information transfer (or filtering) in a wavelength independent fashion[35]. As a corollary, the wavelength and even spectral bandwidth of signals in the two input ports can vary drastically, with little or no impact on the information transduction performance. This spectral insensitivity is a tremendous asset for practical applications since high quality lasers are no longer essential for many RF-photonic applications.

Such wavelength insensitivity, combined with the high power handling and dynamic range, contrasts sharply with the properties of widely studied resonant-cavity based systems for narrow-band signal processing in silicon[4,34,45,48–51]. In addition, this hybrid photonic-phononic emit-receive approach yields filter shape and frequency that is unaltered as optical power is varied by orders of magnitude. This approach negates the need for frequency stabilization (or frequency locking), which often limits the practical utility of resonant optical filtering[4,34,45,48–51]. More generally, since this emit-receive concept opens up a large design space for hybrid photonic-phononic design, it is the impetus for numerous powerful new signal processing schemes.

**Methods:**
**Fabrication Methods**
The silicon cores of the PnC-BAM waveguides were patterned on a silicon-on-insulator wafer with a 3000-nm oxide undercladding using an ASML deep UV scanner, and etched in an AMAT DPS polysilicon etch tool. Following the resist strip and standard post-etch and pre-diffusion cleans, a 300nm LPCVD $Si_3N_4$ was deposited in an SVG series 6000 vertical furnace at about 800 °C. A chemical-mechanical polish was used to preferentially thin down the conformal nitride atop the silicon cores. A hot phosphoric acid etch was used to clear the remaining nitride atop the silicon waveguide. The net result is the waveguide cross-section seen in Fig. 2c. The wafer was patterned again, and then the nitride layer was etched to form the photonic crystals seen in Fig. 2b. Facets for fiber access were obtained by patterning resist with a 1x mask in a SUSS MA-6 contact aligner and etching through deep-RIE process. The oxide under-cladding was then removed with a 49% HF etch.

**Experimental Methods**
The pump (1547 nm) beam is modulated using a Mach-Zehnder intensity modulator with the carrier-frequency component suppressed by a proper bias voltage. The probe (1536 nm) beam split into two paths to form a heterodyne interferometer, where the upper path interrogates the device under test (DUT) and the lower path serves as the local oscillator. The probe beam in the upper path and the pump beam are coupled separately into the two channels of the DUT using two lensed fibres. The probe beam is then coupled out of the DUT using a lensed fibre, where stray pump light (if any) is filtered out with an interference filter. The probe beam in lower path passes through an acousto-optic modulator with a frequency shift ($\Delta/2\pi$) of 40 MHz, and is combined with the probe beam in the upper path using a directional coupler. The beat note between local oscillator ($\omega_3 + \Delta$) and the scattered light from the DUT ($\omega_3 + \Omega$) is recorded using a high speed photodetector, and is analyzed with an RF spectrum analyser. An estimated fibre-to-chip coupling loss of 15 dB, and waveguide propagation loss of < 1 dB/cm were found through waveguide cutback measurements. Pump and probe powers internal to the waveguide are estimated to be 7 mW and 6.3 mW respectively.

**References**
1. Ludwig, M., Safavi-Naeini, A. H., Painter, O. & Marquardt, F. Enhanced Quantum Nonlinearities in a Two-Mode Optomechanical System. *Phys. Rev. Lett.* **109**, 063601 (2012).
2. Brooks, D. W. C. *et al.* Non-classical light generated by quantum-noise-driven cavity optomechanics. *Nature* **488**, 476–80 (2012).
3. Van Thourhout, D. & Roels, J. Optomechanical device actuation through the optical gradient force. *Nat. Photonics* **4**, 211–217 (2010).
4. Hill, J. T., Safavi-Naeini, A. H., Chan, J. & Painter, O. Coherent optical wavelength conversion via cavity optomechanics. *Nat. Commun.* **3**, 1196 (2012).
5. Safavi-Naeini, A. H. *et al.* Electromagnetically induced transparency and slow light with optomechanics. *Nature* **472**, 69–73 (2011).
6. Bochmann, J., Vainsencher, A., Awschalom, D. D. & Cleland, A. N. Nanomechanical coupling between microwave and optical photons. *Nat. Phys.* **9**, 712–716 (2013).
7. Tallur, S. & Bhave, S. A. a silicon electromechanical photodetector. *Nano Lett.* **13**, 2760–5 (2013).
8. Lee, H. *et al.* Chemically etched ultrahigh-Q wedge-resonator on a silicon chip. *Nat. Photonics* **6**, 369–373 (2012).
9. Byrnes, A. *et al.* Photonic chip based tunable and reconfigurable narrowband microwave photonic filter using stimulated Brillouin scattering. *Opt. Express* **20**, 18836–45 (2012).
10. Gao, F. *et al.* On-chip high sensitivity laser frequency sensing with Brillouin mutually-modulated cross-gain modulation. *Opt. Express* **21**, 8605–8613 (2013).
11. Poulton, C. G. *et al.* Design for broadband on-chip isolator using Stimulated Brillouin Scattering in dispersion-engineered chalcogenide waveguides. *Opt. Express* **20**, 21235–46 (2012).
12. Li, M. *et al.* Harnessing optical forces in integrated photonic circuits. *Nature* **456**, 480–4 (2008).
13. Schliesser, a., Rivière, R., Anetsberger, G., Arcizet, O. & Kippenberg, T. J. Resolved-sideband cooling of a micromechanical oscillator. *Nat. Phys.* **4**, 415–419 (2008).
14. Lin, Q., Rosenberg, J., Jiang, X., Vahala, K. J. & Painter, O. Mechanical oscillation and cooling actuated by the optical gradient force. *Phys. Rev. Lett.* **103**, 103601 (2009).
15. Bahl, G., Tomes, M., Marquardt, F. & Carmon, T. Observation of spontaneous Brillouin cooling. *Nat. Phys.* **8**, 203–207 (2012).
16. Chan, J. *et al.* Laser cooling of a nanomechanical oscillator into its quantum ground state. *Nature* **478**, 89–92 (2011).
17. Zhang, J., Peng, K. & Braunstein, S. Quantum-state transfer from light to macroscopic oscillators. *Phys. Rev. A* **68**, 013808 (2003).
18. Palomaki, T. a., Harlow, J. W., Teufel, J. D., Simmonds, R. W. & Lehnert, K. W. Coherent state transfer between itinerant microwave fields and a mechanical oscillator. *Nature* **495**, 210–4 (2013).
19. Kippenberg, T. J. & Vahala, K. J. Cavity opto-mechanics. *Opt. Express* **15**, 17172–205 (2007).
20. Kippenberg, T. J. & Vahala, K. J. Cavity optomechanics: back-action at the mesoscale. *Science* **321**, 1172–6 (2008).
21. Fiore, V., Dong, C., Kuzyk, M. & Wang, H. Optomechanical light storage in a silica microresonator. *Phys. Rev. A* **87**, 023812 (2013).
22. Zhu, Z., Gauthier, D. J. & Boyd, R. W. Stored light in an optical fiber via stimulated Brillouin scattering. *Science* **318**, 1748–50 (2007).
23. Li, J., Lee, H. & Vahala, K. J. Microwave synthesizer using an on-chip Brillouin oscillator. *Nat. Commun.* **4**, 2097 (2013).
24. Marpaung, D., Morrison, B., Pant, R. & Eggleton, B. J. Frequency agile microwave photonic notch filter with anomalously high stopband rejection. *Opt. Lett.* **38**, 4300–3 (2013).
25. Pant, R. *et al.* On-chip stimulated Brillouin Scattering for microwave signal processing and generation. *Laser Photon. Rev.* **14**, n/a–n/a (2014).
26. Marpaung, D., Pagani, M., Morrison, B. & Eggleton, B. Nonlinear Integrated Microwave Photonics. *J. Light. Technol.* **8724**, 1–1 (2014).
27. Eggleton, B. J., Poulton, C. G. & Pant, R. Inducing and harnessing stimulated Brillouin scattering in photonic integrated circuits. 536–587 (2013). doi:10.1364/AOP
28. Vahala, K. *et al.* A phonon laser. *Nat. Phys.* **5**, 682–686 (2009).
29. Grudinin, I. S., Lee, H., Painter, O. & Vahala, K. J. Phonon Laser Action in a Tunable Two-Level System. *Phys. Rev. Lett.* **104**, 083901 (2010).
30. Safavi-Naeini, A. H. & Painter, O. Proposal for an optomechanical traveling wave phonon–photon translator. *New J. Phys.* **13**, 013017 (2011).
31. Zheng, J. *et al.* Feedback and harmonic locking of slot-type optomechanical oscillators to external low-noise reference clocks. *Appl. Phys. Lett.* **102**, 141117 (2013).
32. Marpaung, D. *et al.* Integrated microwave photonics. *Laser Photon. Rev.* **7**, 506–538 (2013).
33. Seeds, A. J. & Williams, K. J. Microwave Photonics. *J. Light. Technol.* **24**, 4628–4641 (2006).
34. Li, H., Chen, Y., Noh, J., Tadesse, S. & Li, M. Multichannel cavity optomechanics for all-optical amplification of radio frequency signals. *Nat. Commun.* **3**, 1091 (2012).
35. Shin, H. *et al.* Tailorable stimulated Brillouin scattering in nanoscale silicon waveguides. *Nat. Commun.* **4**, 1944 (2013).
36. Lin, Q., Painter, O. J. & Agrawal, G. P. Nonlinear optical phenomena in silicon waveguides: modeling and applications. *Opt. Express* **15**, 16604–44 (2007).
37. Rakich, P., Reinke, C., Camacho, R., Davids, P. & Wang, Z. Giant Enhancement of Stimulated Brillouin Scattering in the Subwavelength Limit. *Phys. Rev. X* **2**, 011008 (2012).
38. Russell, P. S. J., Culverhouse, D. & Farahi, F. Experimental observation of forward stimulated Brillouin scattering in dual-mode single-core fibre. *Electron. Lett.* **26**, 1195–1196 (1990).
39. Kang, M. S., Nazarkin, a., Brenn, a. & Russell, P. S. J. Tightly trapped acoustic phonons in photonic crystal fibres as highly nonlinear artificial Raman oscillators. *Nat. Phys.* **5**, 276–280 (2009).
40. Qiu, W. *et al.* Stimulated Brillouin scattering in nanoscale silicon step-index waveguides: a general framework of selection rules and calculating SBS gain. *Opt. Express* **21**, 31402 (2013).
41. Khelif, a., Aoubiza, B., Mohammadi, S., Adibi, a. & Laude, V. Complete band gaps in two-dimensional phononic crystal slabs. *Phys. Rev. E* **74**, 046610 (2006).
42. Haus H A. *Waves And Fields In Optoelectronics*. (Prentice-Hall, 1984).
43. Little, B. E., Chu, S. T., Haus, H. a., Foresi, J. & Laine, J.-P. Microring resonator channel dropping filters. *J. Light. Technol.* **15**, 998–1005 (1997).

44. Little, B. E. *et al.* Very High-Order Microring Resonator Filters for WDM Applications. *IEEE Photonics Technol. Lett.* **16,** 2263–2265 (2004).
45. Chan, E. H. W., Alameh, K. E. & Minasian, R. A. Photonic Bandpass Filters With High Skirt Selectivity and Stopband Attenuation. *J. Light. Technol.* **20,** 1962–1967 (2002).
46. Raymer, M. G. & Srinivasan, K. Manipulating the color and shape of single photons. *Phys. Today* **65,** 32–37 (2012).
47. Monroe, C. Quantum information processing with atoms and photons. *Nature* **416,** 238–46 (2002).
48. Capmany, J., Member, S. & Ortega, B. A Tutorial on Microwave Photonic Filters. **24,** 201–229 (2006).
49. Chan, E. H. W., Minasian, R. A. & Abstract, A. Coherence-Free High-Resolution RF / Microwave Photonic Bandpass Filter With High Skirt Selectivity and High Stopband Attenuation. *J. Light. Technol.* 1646–1651 (2010).
50. Alipour, P. *et al.* Fully reconfigurable compact RF photonic filters using high-Q silicon microdisk resonators. *Opt. Express* **19,** 15899–907 (2011).
51. Supradeepa, V. R. *et al.* Comb-based radiofrequency photonic filters with rapid tunability and high selectivity. *Nat. Photonics* **6,** 186–194 (2012).

Acknowledgements: Sandia Laboratory is operated by Sandia Co., a Lockheed Martin Company, for the U.S. Department of Energy's NNSA under Contract No. DE-AC04-94AL85000. This work was supported in part by the DDRE under Air Force Contract No. FA8721-05-C-000, the MesoDynamic Architectures program at DARPA under the direction of Dr. Jeffrey L. Rogers and Dr. Daniel Green, and Sandia's Laboratory Directed Research and Development program under Dr. Wahid Hermina. Z.W. acknowledges support from the Packard Fellowship in Science and Engineering and the Alfred P. Sloan Research Fellowship. We thank Dr. Wenjun Qiu, Dr. Ryan M. Camacho, Dr. Roy Olsson, and Dr. Ihab El-Kady for helpful technical discussions involving phononic systems, optomechanics and nonlinear interactions. We are grateful to Dr. Ryan Behunin, Dr. William Renninger, Dr. Whitney Purvis Rakich for careful reading and critique of this manuscript.

# Supplementary Information for "Control of coherent information via on chip photon-phonon emitter-receiver"

## Supplementary Note 1

### Wavelength and laser linewidth insensitivity.

The generation and detection of coherent phononic waves are proportional to the magnitude of the optical power modulation[1]. This indicates that laser linewidth of the waves in either optical waveguides does not limit the performance of the PPER system. For instance, the laser linewidth of the probe beam under our experimental conditions is about 5 MHz, however we observed much narrower spectrum features than the laser linewidth. In addition, the use of traveling-wave schemes requires that the group-velocity mismatch between two optical waves in either port should be less than the signal modulation period to reduce the temporal walk-off and to achieve effective photon-phonon coupling.

$$(n_{g,A} - n_{g,B})L < 2\pi c/\omega_0 n_{g,A}, \tag{1}$$

where $n_{g,j}$ represents the group index in waveguide $j$, $L$ is the length of waveguides, $c$ is the speed of light, and $\omega_0$ indicates the resonant frequency of a PPER system. The group index and effective refractive index of the waveguide are calculated from its geometry as in Fig. S1 using 2-D vector-field modesolver.

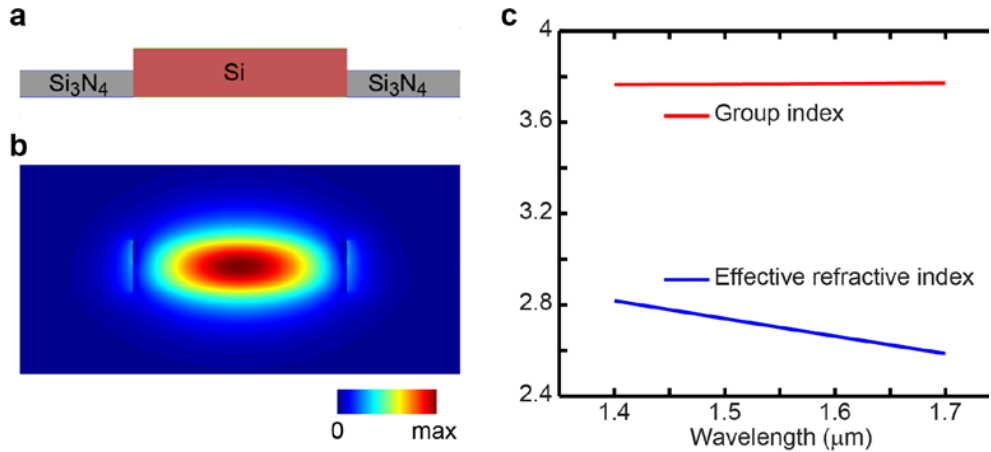

**Figure S1| Group index and effective refractive index of the optical waveguides. a,** The waveguide geometry used in the simulation. The silicon waveguide core has the size of $950 \times 220 \text{ nm}^2$, and the thickness of silicon nitride membrane is 130 nm. **b,** Computed $E_x$ fields of the guided optical mode in the silicon waveguide. **c,** Computationally calculated group index and effective refractive index of the waveguide.

As seen in Fig. S1c, the group index variance of the optical waveguides ($950 \times 220 \text{ nm}^2$) is less than 0.2 % over 1.4 μm to 1.7 μm while the effective refractive index changes by about 9%. This result indicates that the information transfer is largely insensitive over wide range of wavelength with the typical length-scale of the optical waveguides. From these results, we believe that even incoherent light sources (e.g. light emitting diode) can be used which could be a tremendous asset of the PPER system for practical applications in optical signal processing.

## Supplementary Note 2

### Power handling capability.

We theoretically predict the power handling capability of the silicon waveguide using its propagation loss. In this article, we consider the propagation loss contributed by linear absorption (material absorption and scattering effects) and nonlinear absorption (two photon absorption (TPA) and TPA induced free-carrier absorption) which has been well studied in silicon[2,3]. Then the total absorption coefficient, $\alpha$, can be given by

$$\alpha = \alpha_o + \beta_{TPA} \frac{P}{A} + \sigma \Delta N \left(\frac{P}{A}\right)^2, \tag{S2a}$$

where $\alpha_o$ and $\beta_{TPA}$ represent the linear absorption and TPA coefficients, respectively. $P$ is the internal optical power in the waveguide, and $A$ is the effective mode area of silicon waveguide. $\sigma$ represents the free-carrier absorption cross-section, and $\Delta N = \tau \beta_{TPA}/2h\nu$ is the free-carrier number density in steady state conditions. $\tau$ is the free-carrier lifetime in silicon waveguide, and $h\nu = 1.28 \times 10^{-19}$ [ J ] is the photon energy of the pump light at 1550 nm. The values of $\beta_{TPA} = 0.5 \times 10^{-11}$ [m/W] and $\sigma = 1.45 \times 10^{-21}$ [m$^2$] in silicon are well studied[2,3]. The effective mode area, $A = 1.3 \times 10^{-13}$ [m$^2$], is numerically solved using COMSOL multi-physics simulations using the physical size of silicon waveguide of $950 \times 220$ nm$^2$. We measured the free-carrier lifetime of $\tau = 2$ ns in our silicon waveguide with a pump-probe experiment, and the linear absorption coefficient, $\alpha_o = 1$ [dB/cm] = 23 [m$^{-1}$], is extracted from a cut-back measurement. Using the parameters above, the total absorption coefficient of Eq. S1a in the silicon waveguide can be rewritten as,

$$\alpha = 23 + 38.46 \times P + 3352 \times P^2 \ [m^{-1}]. \tag{S2b}$$

In this article, we define the power handling capability of the silicon waveguide as the power yielding the total absorption coefficient of $\alpha = 3$ dB/cm = 69 [m$^{-1}$], which was chosen to be consistent with the length-scale of propagation on silicon chips. Hence, the internal optical power of $P = 110$ mW is the power handling of our silicon waveguide. This power in the silicon waveguides can be easily achievable with better fiber-to-waveguide coupling method.

## Supplementary Note 3

### Temporal coupled mode theory.

The analytical expression of the photon-phonon emitting-receiving functionalities can be achieved using temporal coupled-mode theory[4,5]. As discussed in the main context, we consider two pump fields $E_1^a(\omega_1)$ and $E_2^a(\omega_2)$ in Wg-A and probe as well as signal fields $E_3^b(\omega_3)$ and $E_s^b(\omega_s)$ in Wg-B. Here, $E_j^a(\omega_j, t) \equiv h_a(x, y)A_j(z)e^{i(k_j z - \omega_j t)}$ and $E_j^b(\omega_j) \equiv h_b(x, y)B_j(z)e^{i(k_j z - \omega_j t)}$, where $h_a$ ($h_b$) represents the field mode distribution in Wg-A (Wg-B), and $A_j$ and $B_j$ are power-normalized mode amplitudes such that $|A_j|^2$ ($|B_j|^2$) is the modal power, $P_j^a$ ($P_j^b$) carried in Wg-A (Wg-B).

By the existence of optical fields in waveguide, the optical forces mediated by electrostriction and radiation pressure can yield elastic displacement within the silicon core. The optical force induced by an optical wave $\boldsymbol{E}_j^a(\omega_j)$ can be written as, $\mathbf{f_a} = \mathbf{f}_n^a(x,y)\left|A_j\right|^2 = \mathbf{f}_n^a(x,y)\,P_j^a$, where $\mathbf{f}_n^a(x,y)$ is the force density normalized to the modal power in Wg-A under continuous-wave excitation. The optical force density distribution induced by two optical fields ($\boldsymbol{E}_1^a(\omega_1)$ and $\boldsymbol{E}_2^a(\omega_2)$) in Wg-A consists of a constant dc term and an oscillating term as, $\mathbf{f_a} = \mathbf{f_{DC}} + \mathbf{f_{AC}} = \mathbf{f}_n^a(x,y)(|A_1|^2 + |A_2|^2) + \mathbf{f}_n^a(x,y)2A_1A_2^* e^{i(Kz-\Omega t)}$, where $\Omega = (\omega_1 - \omega_2)$ is the beating frequency and $\mathbf{K} = \mathbf{k_1} - \mathbf{k_2}$ is the phonon wave-vector. Note that we ignore the phase difference between two optical fields.

The oscillating optical force of $\mathbf{f}_n^a(x,y)2A_1A_2^* e^{-i\Omega t}$ yields elastic displacement field, $\boldsymbol{u_a} = \boldsymbol{e_a}(x,y)c_a(t) = \boldsymbol{e_a}(x,y)C_a e^{-i\Omega t}$, in Wg-A. Note that the optical cross-talk between Wg-A and Wg-B is negligible, but phonons can transfer to Wg-B through the central PnC region, creating phononic defect mode in Wg-B, $\boldsymbol{u_b} = \boldsymbol{e_b}(x,y)c_b(t) = \boldsymbol{e_b}(x,y)C_b e^{-i\Omega t}$. In the limit of small signal amplitudes, $B_s$, the phonon dynamics at any position ($z$) along the length of the waveguides can be expressed as,

$$\frac{dc_a(t)}{dt} = -\left(i\Omega_0 + \frac{1}{\tau_{net}}\right)c_a(t) + i\mu c_b(t) + \eta(t)A_1A_2^*, \tag{S3a}$$

$$\frac{dc_b(t)}{dt} = -\left(i\Omega_0 + \frac{1}{\tau_{net}}\right)c_b(t) + i\mu c_a(t), \tag{S3b}$$

using temporal coupled-mode theory[4,5]. Here, $\Omega_0$ is the natural frequency of the uncoupled phonon modes, $\mu$ represents the phononic modal coupling rate, and $\tau_{net}^{-1}$ is the net decay rate of phonon mode, which is related to the external decay ($\tau_e^{-1}$) through either side of PnC cladding and internal ($\tau_o^{-1}$) decay rates as $\tau_{net}^{-1} = \tau_e^{-1} + \tau_o^{-1}$. Above, $\eta(t)A_1A_2^*$ is the driving term, of the form $\eta(t) = \eta_o\exp(-i\Omega t)$. From Ref. [6], the coupling amplitude, $\eta_o$, becomes $\eta_o = \langle \mathbf{f}_n^a, \boldsymbol{e_a}\rangle \times \langle \boldsymbol{e_a}, \rho\boldsymbol{e_a}\rangle^{-1}(\Omega_o)^{-1}$, where $\langle \boldsymbol{X}, \boldsymbol{Y}\rangle \equiv \int \boldsymbol{X}^* \cdot \boldsymbol{Y} dS$ over the waveguide cross-section and $\rho(x,y)$ is the mass density of the elastic medium. Solutions of Eq. S2a and Eq. S2b yield a set of hybridised (symmetric and anti-symmetric) phonon modes with wave amplitudes $C_a = i\eta_0(\Omega - \Omega_o + i/\tau_{net})A_1A_2^*\left[\Gamma_-(\Omega)\,\Gamma_+(\Omega)\right]^{-1}$, and $C_b = i\mu\eta_0A_1A_2^*\left[\Gamma_-(\Omega)\,\Gamma_+(\Omega)\right]^{-1}$, where $\Gamma_{\pm}(\Omega) \equiv [\Omega - (\Omega_0 \pm \mu) + i/\tau_{net}]$. Hence, optical wave-mixing in Wg-A drives an elastic displacement field ($\boldsymbol{u_b}(\boldsymbol{t})$) in Wg-B.

Now we consider the impact of elastic deformation induced by the phonon field on optical waves propagating in Wg-B. Since light guided in the core of the waveguide produces a force distribution within the core of the waveguide, any elastic displacements will perform work against these optical forces, changing the energy of the guided electromagnetic fields. Using the principle of virtual work, the change in guided electromagnetic energy per unit length in Wg-B produced by a small elastic deformation of amplitude $\delta\boldsymbol{u_b}$ can be expressed as $\delta U_{ME}/L = \langle \delta\boldsymbol{u_b}, \mathbf{f}^b\rangle = \langle \boldsymbol{e_b}, \mathbf{f}_n^b\rangle\delta c_b P_3^b$. Following Refs. [1,7], one can show that this change in electromagnetic energy is equivalent $\delta U_{EM} = (P_3^b/\omega_3)\,\delta\phi$ where $\delta\phi$ represents the phase induced by variation in the mechanical degree of freedom $\delta c_b$. Hence, the phase change per unit length imparted on wave $\tilde{B}_3$ by elastic wave displacement, $\boldsymbol{u_b}(\boldsymbol{t})$, can be expressed as $\delta\phi/L = (1/L)(\delta\phi/\delta c_b)C_b = \langle \boldsymbol{e_b}, \mathbf{f}_n^b\rangle C_b\omega_3$. Using this result, we have the phonon mediated coherent coupling from Wg-A to Wg-B, $\gamma_{a\rightarrow b}(\Omega)$, as

$$\gamma_{a\to b}(\Omega) = \left[ \frac{\omega_3 \tau_{\text{net}}}{2\Omega_0} \frac{\langle \mathbf{f}_n^a, \boldsymbol{e}_a \rangle \langle \boldsymbol{e}_b, \mathbf{f}_n^b \rangle}{\langle \boldsymbol{e}_a, \rho \boldsymbol{e}_a \rangle} \frac{2\mu/\tau_{\text{net}}}{\left[ \Omega - \Omega_0 - \sqrt{\mu^2 - 1/\tau_{\text{net}}^2} \right] \left[ \Omega - \Omega_0 + \sqrt{\mu^2 - 1/\tau_{\text{net}}^2} \right]} \right]. \tag{S3c}$$

Then phonon mediated signal field change per unit length in Wg-B can be rewritten as,

$$\frac{\partial B_{\text{S}}}{\partial z} = i[|\gamma_{a\to b}(\Omega)] A_1^* A_2 B_3 \tag{S3d}$$

In low gain regime, the output signal power, $P_s^b$, is given by $P_s^b = |\gamma_{a\to b}(\Omega)|^2 P_1^a \ P_2^a \ P_3^b \ L^2$.

## Supplementary Data

## Spectra of photon-phonon emitter-receiver responses over the stopband frequency range.

We also investigate the responses of the photon-phonon emitter-receiver systems over the stopband frequency range (2.6 – 4.5 GHz, yellow region). The dual channel PnC-BAM waveguides for $W_o = 5.7$ μm and $W_o = 5.2$ μm are tested under identical experimental conditions. As seen in Fig. S1, the normalized RF power for $W_o = 5.7$ μm ($W_o = 5.2$ μm ) shows a strong peak at 2.93 GHz (3.18 GHz) and flat response over the stopband range. The response for $W_o = 5.7$ μm displays a narrow feature at 4.44 GHz which is a higher order resonant mode. Several signatures produced by complicated phononic modes are also shown at around 3.7 GHz for $W_o = 5.7$ μm, but not for $W_o = 5.2$ μm, indicating that these complicated features can be removed by engineering the phononic supermodes and the PnC structures.

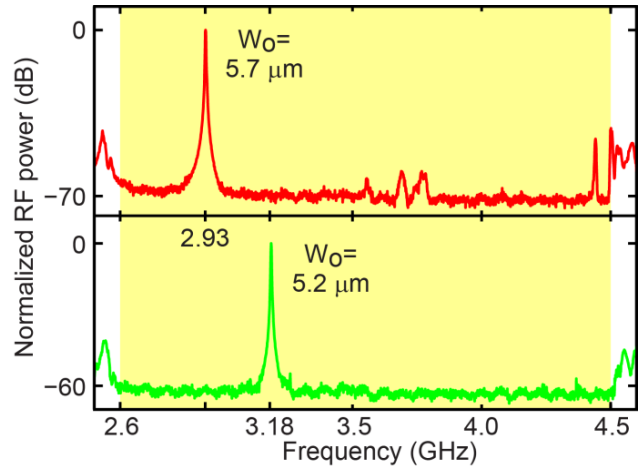

**Figure S2| Spectra of photon-phonon emitter-receiver functionalities over the stopband range.** The normalized RF responses produced by the signal field are displayed through the dual channel PnC-BAM waveguides for $W_o = \mathbf{5.7}$ **μm** (red) and $W_o = \mathbf{5.2}$ **μm** (green).

## Supplementary references

1. Rakich, P., Reinke, C., Camacho, R., Davids, P. & Wang, Z. Giant Enhancement of Stimulated Brillouin Scattering in the Subwavelength Limit. *Phys. Rev. X* **2,** 011008 (2012).

2. Lin, Q., Painter, O. J. & Agrawal, G. P. Nonlinear optical phenomena in silicon waveguides: modeling and applications. *Opt. Express* **15,** 16604–44 (2007).

3.    Lin, Q. *et al.* Dispersion of silicon nonlinearities in the near infrared region. *Appl. Phys. Lett.* **91,** 021111 (2007).

4.    Haus H A. *Waves And Fields In Optoelectronics*. (Prentice-Hall, 1984).

5.    Little, B. E., Chu, S. T., Haus, H. a., Foresi, J. & Laine, J.-P. Microring resonator channel dropping filters. *J. Light. Technol.* **15,** 998–1005 (1997).

6.    Qiu, W. *et al.* Stimulated Brillouin scattering in nanoscale silicon step-index waveguides: a general framework of selection rules and calculating SBS gain. *Opt. Express* **21,** 31402 (2013).

7.    Rakich, P. T., Popović, M. a & Wang, Z. General treatment of optical forces and potentials in mechanically variable photonic systems. *Opt. Express* **17,** 18116–35 (2009).